\documentclass[]{article}
\usepackage{proceed2e}

\usepackage{times}
\usepackage[pdftex]{graphicx} 

\usepackage{qiangstyle}
\usepackage{times}

\title{Taxable Stock Trading with Deep Reinforcement Learning}

\author{
{\bf Shan Huang} \\
National University of Singapore,
\texttt{a0120756@u.nus.edu}
}

\begin{document}

\maketitle

\begin{abstract}
 In this paper, we propose stock trading based on the average tax basis. Recall that when selling stocks, capital gain should be taxed while capital loss can earn certain tax rebate. 
We learn the optimal trading strategies with and without considering taxes by reinforcement learning. The result shows that tax ignorance could induce more than 62\% loss on the average portfolio returns, implying that taxes should be embedded in the environment of continuous stock trading on AI platforms.  
\end{abstract}

\section{Background}
Stochastic control over time is one of the most important topics in financial trading, portfolio selection, and asset allocation. In stochastic control,  an agent optimally makes the decision (action) based on the observed state variables, in order to maximize his objective function. 
Recent years have seen a lot of successful applications of deep reinforcement learning to train a self learning AI-Agents. For example, AlphaGo, a hybrid DRL system, defeated a human world champion in Go \citep{S16}.  DRL algorithms have already been applied to a wide range of problems, such as robotics \citep{L16, D16} and video games \citep{M15}. Moreover, several different approaches have been proposed for reinforcement learning with neural network function approximators \citep{M15, SM15, SM16, SM17}
 Trading stocks by reinforcement learning can guide and  help agents to increase their portfolio returns. Though transaction cost is considered when buying and selling stocks\footnote{See \url{https://github.com/hackthemarket/gym-trading}}, those results are still questionable since tax is never considered. 
Paying tax should be the one of the main concerns in stock trading because tax is much higher compared to transaction costs. Notice that reinforcement learning is developed based the Markovian Decision Process (MDP), while the exact tax calculation is not Markovian and thus can not be directly used in reinforcement learning.

Investors in U.S. stock markets are subject to capital gains tax when gains or losses are realized. When gains are realized, a lower long-term tax rate $\alpha_{L} =15\%$ applies if stock holding period is at least one year and a higher short-term tax rate $\alpha_S = 25\%$ applies if stock holding period is less than one year. In contrast, when losses are realized, the investors can get a tax rebate with the short-term rate regardless of the length of the holding period. Tax rebate means that the loss can be deducted from gains and only the remained gains are taxed. 
We use the average-basis and average-holding-time system to simplify the path-dependent tax calculation without affecting main quantitative results. The average basis technique can make the tax calculation Markovian since the average-basis and average-holding-period at current step are updated only using the state variables on the last time step. 
The average tax basis is one of many methods that investors can use to arrive at the cost of their stock holding, mutual fund holdings, and other taxable financial goods. It is known that Australia is the country using average tax basis.  To understand the average-basis and average-holding-period scheme, we give an example as follows. Assume that the investor bought 300 shares of stock SPY at price $\$200$ per share two years ago and purchased 100 more shares at $\$300$ per share half a year ago. Now he sells the total 400 shares at price $\$350$ per share. The total cost basis equals $\$200  \times 300$ $+ \$ 300\times 100 = \$90,000$ and the
average basis is $\$90,000/(300+100) = \$225$ per share. The basis-weighted total holding time
is $\$200 \times 300 \times 2 + \$300\times 100\times 0.5 = 135,000$ (dollar year) and 
 the average holding period equals $ 135,000 /90,000 = 1.5$ years. In this way,  the total capital gain after selling equals $(\$350\times 400 - \$225 \times 400) = \$50,000$. Since the average holding period is above one year, the capital gain should be taxed at the long-term rate and thus, the tax being charged equals $\$50,000\times 15\% = \$7,500$. Compared to taxes, transaction cost per trading is only around $0.1\%\sim 0.5\%$. Therefore the maximum transaction costs involved equal $(\$200  \times 300 \times 0.005+\$300\times 100\times 0.005 + \$350\times 400\times 0.005) = \$1,150$, which is much lower than the tax charged. This example demonstrates the indispensability of tax consideration in stock trading. A continuous stochastic dynamic model can be found in \cite{D15}. 

\section{Model}
Taking the stock price $s_t$, average-basis $b_t$, and average-holding-time $h_t$ as state variables, the stock trading problem becomes a MDP problem and we can program an AI-agent with reinforcement learning.  
The policy gradient method for reinforcement learning works by computing an estimator of the policy gradient and plugging it into a stochastic gradient ascent algorithm:
\begin{equation}
    g = {\mathbb{E}} \Big[ \sum_{t=0}^T A_t \nabla_{\theta} \log \pi_{\theta}(a_t | s_t, b_t, h_t) \Big],
\end{equation}
where $a_t$ is the action of stock trading following $a_t \sim \pi_\theta(a_t | s_t, b_t, h_t)$ and $ A_t$ is an estimator of the advantage function at timestep $t$,   The advantage function $A^{\pi}(s,b,h,a) = Q^{\pi}(s,b,h, a) -V^{\pi}(s,b,h)$, where 
$$Q^{\pi}(s,b,h, a) = \mathbb{E}_{\pi}\Big[\sum_{l=0}^{\infty} \gamma^{l} r_{t+l} | s_t = s, b_t = b, h_t =h, a_t =a\Big]$$ and $$V^{\pi}(s,b,h) =  \mathbb{E}_{\pi}\Big[\sum_{l=0}^{\infty} \gamma^{l} r_{t+l} | s_t = s, b_t = b, h_t =h\Big]$$ with $\gamma$ the discounted factor and $r_t$ the reward at timestep $t$. 
Notice that the advantage function
measures whether or not the action is better or worse than the policy's default behavior. The multiplication of $A_t$ and $\nabla_{\theta}\log \pi_{\theta}$ implies that a step in the policy gradient direction should increase the probability of better-than-average actions and decrease the probability of worse than average actions. We choose proximal policy optimization algorithms (PPO) in Schulman et al. (2017), which outperforms other online policy gradient methods, and overall strikes a favorable balance between sample
complexity, simplicity, and wall-time.  

We can show that the state process $(s_t,b_t,h_t)$ are Markovian satifying $s_t,b_t,h_t$  $\geq 0$. Recall the definition of average-basis $b$ and average-holding-period $h$ for taxes. 
The evolution of $b$ and $h$ depends on action $a$ and the observed stock price $s$. If we denote by $a_t$ the the shares of stock holdings at timestep $t$, then the average-basis at next timestep is
\begin{equation}\label{averge basis}
    b_{t+1} = \left\{ \begin{aligned}
    & s_{t+1}  & a_ta_{t+1}\leq 0,\\
    & \frac{b_t a_t +s_{t+1}(a_{t+1}-a_{t})}{a_{t+1}} & a_{t+1}<a_{t} < 0, \\
    & \frac{b_{t} a_{t} +s_{t+1}(a_{t+1}-a_{t})^+}{\max(a_{t},a_{t+1})} & \text{otherwise}, \\
\end{aligned}\right.
\end{equation} 
where $s_{t+1}$ and $a_{t+1}$ are the stock price and stock positions at timestep $t+1$ respectively. 
The update of average-basis depends on the relation among $a_t$, $0$, and $a_{t+1}$. All history basis record will be waived when stock position goes across $0$. For example, if $a_ta_{t+1}\leq 0$, the stock position changes from short to long, or long to short, the average-basis is reset to be $s_{t+1}$ since all history transactions are finished. When $a_{t+1}<a_{t}<0$, the agent decides to continue shorting stocks so that the average-basis of shorting is the total cost basis $b_t a_t + s_{t+1}(a_{t+1}-a_{t})$ (negative value) divided by current position $a_{t+1}$ (negative value). For other cases, stock buying can change the average-basis by varying total cost basis and stock holdings differently, while stock selling does not change the average-basis because stock selling proportionally decreases the total cost basis and stock positions. That gives the last equality in (\ref{averge basis}). We have embedded short selling into our average basis system.  In finance, short selling is the sale of a security that the seller has borrowed.  When shorting stocks, the investor borrows the shares and immediately sells them. To close the transaction, the investor covers the position by buying the shares later and delivering the securities back to lender. Capital gain is made when the purchase price is lower than the selling price at borrowing and loss is made when the purchase price is higher than the initial selling price. Gain or loss is taxed at the time of the close of transaction. 
Similarly, we can get the average-holding-period at next timestep 
\begin{equation}\label{averge holding period}
    h_{t+1} = \left\{ \begin{aligned}
    & 0 &  a_ta_{t+1}\leq 0, \\
    & \frac{b_t a_t (h_t + dt)}{b_{t+1}a_{t+1}} & a_{t+1}<a_{t} < 0, \\
    & \frac{b_t a_t (h_t + dt)}{b_{t+1}\max(a_{t},a_{t+1})} & \text{otherwise}. \\
\end{aligned}\right.
\end{equation} 

Given the average-basis and average-holding period, we now calculate the tax costs at timestep $t+1$. We first assume that $s_{t+1}\geq b_{t}$. Capital gain is realized when selling stocks or buying stocks to attenuate the previous short exposure. Thus, the capital gain tax at timestep $t+1$ equals 
\begin{eqnarray*}\begin{aligned}
&(s_{t+1} - b_{t})\Big[(a_{t}-a_{t+1}^+) \mathbf{1}_{\{a_t\geq a_{t+1}, a_t\geq 0\}}  \\
&- (a_{t}+a_{t+1}^-) \mathbf{1}_{\{a_t\leq a_{t+1}, a_t\leq 0\}}\Big] \big(\alpha_S \mathbf{1}_{\{h_t<252\}} + \alpha_L \mathbf{1}_{\{h_t\geq 252\}}\big), 
\end{aligned}
\end{eqnarray*}
where $\mathbf{1}_{\{\}}$ is the indicator function. The capital gain is taxed at the long-term rate if the average-holding-period is above one year (252 trading days) and at the short-term rate if the average-holding-period is shorter than one year.
Now we explain the formulas in the bracket above. When investors sell owned stocks at price $s_{t+1}$, it implies that $a_t\geq a_{t+1}\geq 0$. Then the total capital gain equals $(s_{t+1}-b_{t})(a_t-a_{t+1})$. When investors wash sell all the stocks and then continue to short stocks, that is, $a_t\geq 0> a_{t+1}$, the capital gain only comes from washsell and equals $(s_{t+1}-b_{t}) a_t$. Combining these two cases gives the first formula in the bracket above. Alternatively, capital gain can be realized when buying stocks to attenuate the previous short exposure. When investors buy back part of the shorted  stocks, that is, $a_t\leq a_{t+1}\leq 0$, the realized capital gain equals $(s_{t+1}-b_t)(a_{t+1}-a_{t})$. When investors buy back all the shorted stocks and continue to purchase stocks to gain positive exposure, that is, $a_t\leq 0 < a_{t+1}$, the capital gain only comes from the compensation of short selling and equals $(s_{t+1}-b_t)(-a_t)$. Combining these two cases gives the second formula in the bracket above.
 
Similarly, if $s_{t+1}<b_{t}$, there is a capital loss and the tax rebate from capital loss equals
\begin{eqnarray*}\begin{aligned} 
(b_{t} - s_{t+1})\Big[ &(a_{t}-a_{t+1}^+) \mathbf{1}_{\{a_t\geq a_{t+1}, a_t\geq 0\}} \\
&- (a_{t}+a_{t+1}^-) \mathbf{1}_{\{a_t\leq a_{t+1}, a_t\leq 0\}}\Big] \alpha_S.
\end{aligned}\end{eqnarray*}
The tax rate on capital loss is $\alpha_S$ since when losses are realized, the investors get a tax rebate with the short-term rate regardless of the length of the holding period. 
The tax rebate implies that the agent can use capital losses (stock losses) to offset capital gains during a taxable year. If the agent does not have enough capital gains to offset the capital loss,  a capital loss can be used as an offset to ordinary income (assume taxing at short-term tax rate), up to \$3,000 per year. Note that if we do not distinguish the long-term tax and short-term tax, the state variable can be chosen as $(s,b)$ and the average-holding-period is not needed. The tax rebate of capital loss feeds to the reward and the capital gain tax makes a leakage from it. As the capital gain tax and tax rebate affect the reward at each times step, the policy to maximize the expected total reward should be different from those without considering taxes. 



\begin{figure}[h]
\centering
\includegraphics[height=5.5cm,width=9cm]{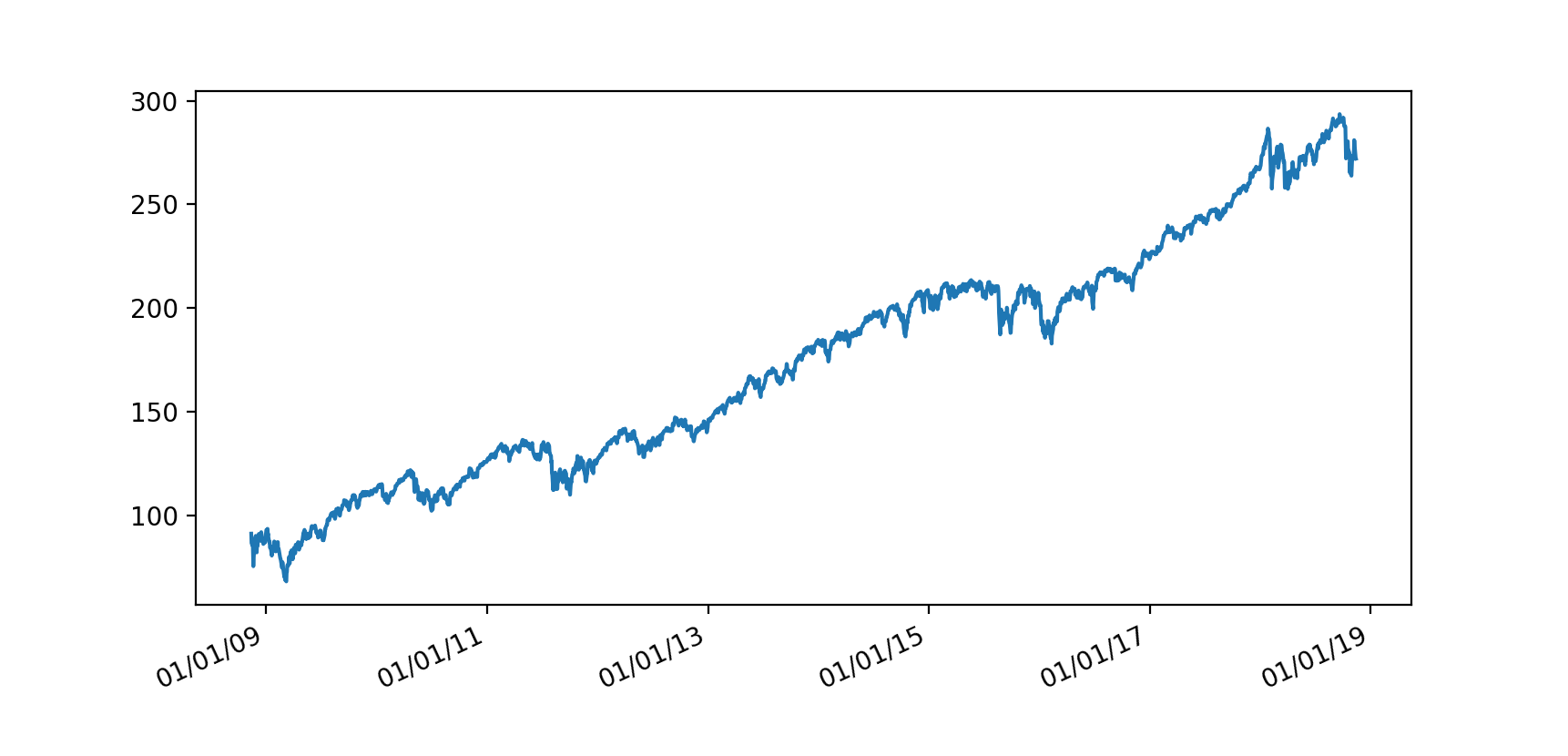}
\caption{SPDR S\&P 500 ETF Trust (SPY) from 13/11/2008 to 13/11/2018. }
\label{Fig1}
\end{figure}

For simplicity, we consider the representative SPY stock trading. Our data set includes SPY's daily closed price and volumes from 13/11/2008 to 13/11/2018, as shown in Fig \ref{Fig1}. We choose the time step $dt =1$ representing one trading day and the total trading days per year is 252 trading days. When the average-holding-period $h$ is larger (smaller) than 252, 15\% (25\%) of the capital gain is charged and $25\%$ of capital loss is rebated. That is, $\alpha_L=0.15\%$ and $\alpha_S=0.25\%$.
Each time the agent can short, long, or not trade stocks. We assume the basic stocks shares per trading is 100 shares and the agent keeps his stock position $-100$, $0$, or $100$ over time. The length of trading period is set to be 5 years, totally $T = 252\times 5 = 1260$ trading days. Transaction cost is also included, which equals $0.1\%$ of the gain or loss per trading.  We create a new OpenAI Gym environment where the observation in each timestep is SPY's daily closed price, trading volume, averaged-basis, and average-holding-period. The action-based evolution of the average basis system is given by (\ref{averge basis}) and (\ref{averge holding period}). 

\begin{figure}[h]
\centering
\includegraphics[height=5cm,width=6cm]{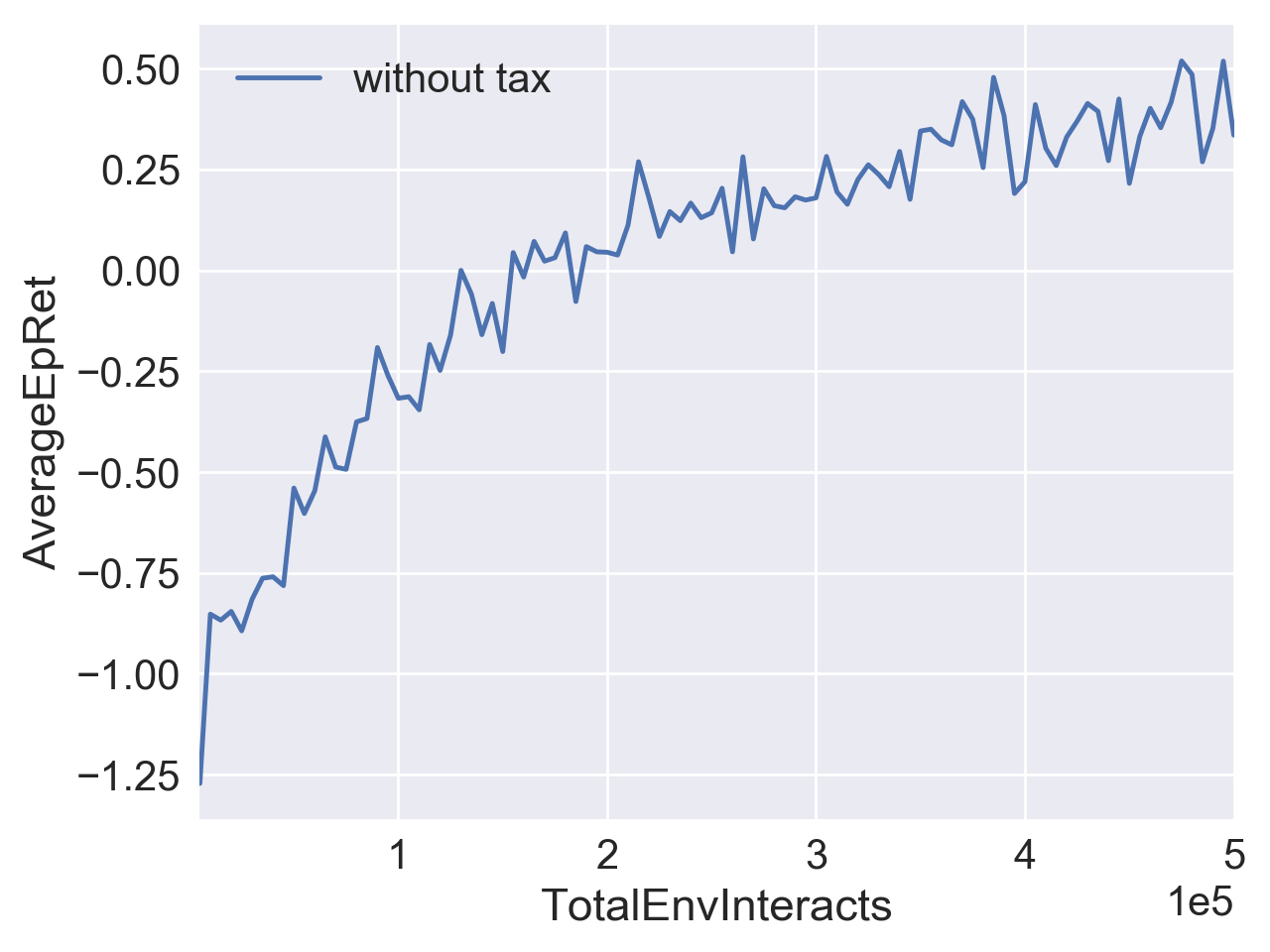}
\includegraphics[height=5cm,width=6cm]{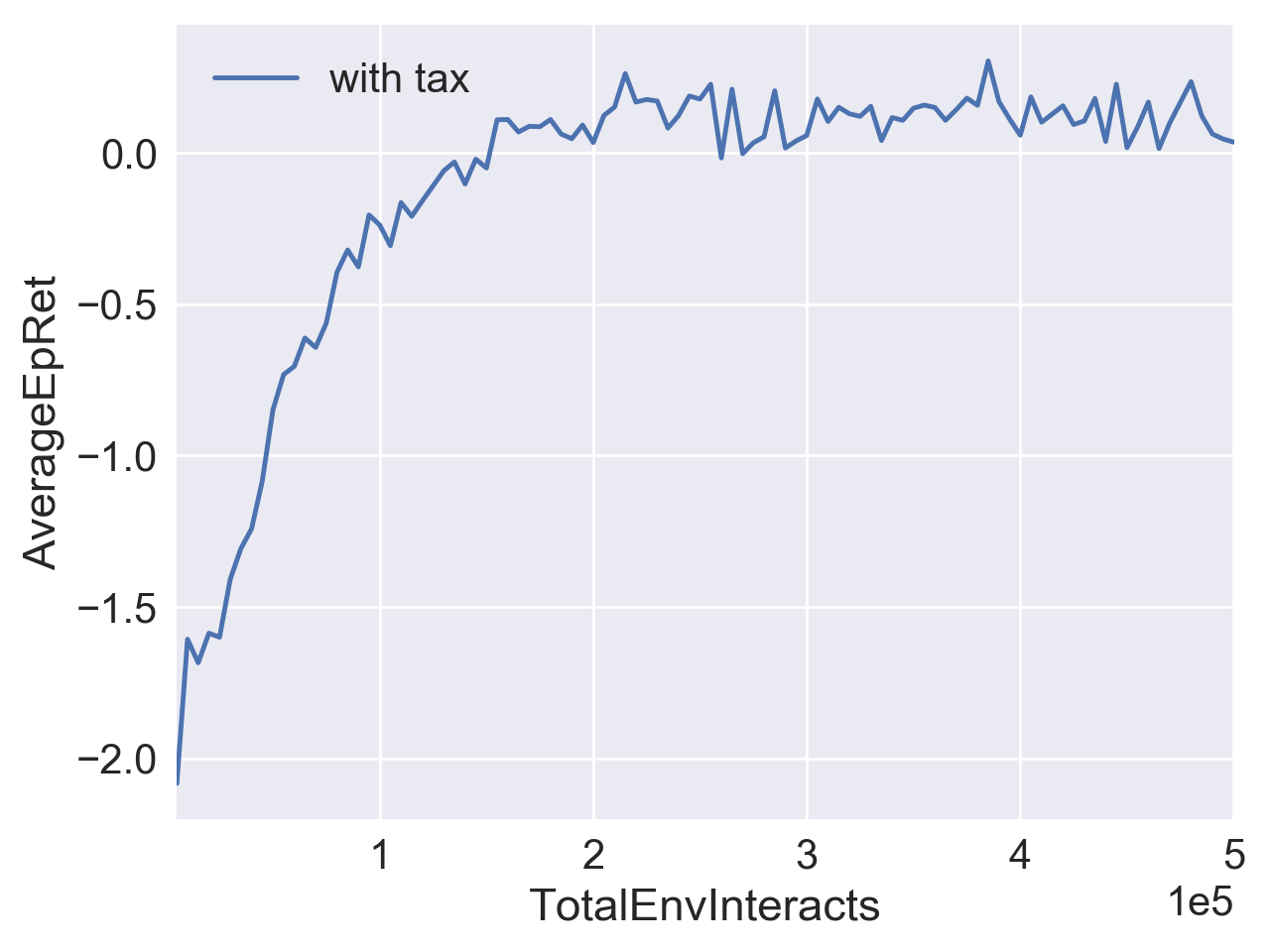}
\caption{Average expected returns on SPY investment. The length of trading period is 5 years with totally $252\times 5 = 1260$ trading days. Long and short tax is considered. When the average-holding-period is larger (smaller) than 252, 15\% (25\%) of the capital gain is charged and $25\%$ of capital loss is rebated. Small transaction cost ($0.1\%$) is also included. }
\label{Fig2}
\end{figure}

To represent the policy, we use the same default neural network architecture as PPO with fixed-length trajectory segments, which was a fully-connected MLP with two hidden layers of 64 and 64 tanh units respectively. 
The final output layer has a linear activation. policy and value function are estimated through separated network. 
The  number of steps of interaction (state-action pairs) for the agent and the environment in each epoch is 5000 and the number of epochs is 50. The hyperparameter for clipping in the policy objective is chosen to be 0.2 and the GAE-Lambda is 0.97. The learning rate for policy and value function optimizer is 0.001 and 0.0003 respectively. If tax is not included in the model, the average expected return is 0.44, as shown in the top panel of Fig \ref{Fig2}, which seems quite promising. This considerable return is the result of exploiting price trending and frequently adjusting holding positions correspondingly, similar as the results of other AI platforms.
However, this is not compelling since tax is heavily charged in a taxable year. Rather than ignoring taxes, the learning of stock trading should consider the effect of tax costs. We use PPO to train the stock trading policy in the environment with tax costs, as shown in the  down panel of Fig \ref{Fig2}. The optimal stock trading policy in the model with taxes can achieve 0.13 average returns. To illustrate the suboptimality of the policy trained in the model without considering taxes (the policy obtained in the top panel of Fig  \ref{Fig2}), 
we apply this trained policy in the environment with tax costs, the average expected return drops to only 0.05. This implies that tax ignorance could induce more than $(0.13-0.05)/0.13 =62\%$ loss on average portfolio returns. In the testing environment, we consider the daily trading, which allows frequent stock holding adjustment. 
 The effect of taxes could be weakened if the time step of trading is chosen to per month or longer.

\section{CONCLUSIONS}
In this paper, we are among the first to embed taxes into reinforcement learning via average basis system. Notice that the exact tax calculation at timestep $t+1$ depends on the whole history path $\{s_i,a_i\}_{i=1,2,..,t}$, which is impossible to be used in reinforcement learning as the state dimension will explode when time period goes long. By introducing the average-basis $b_t$ and the average-holding-time $h_t$, the augmented state variables $(s_t,b_t,h_t)$ becomes Markovian and further the action and tax cos at timestep $t+1$ only depend on  $(s_t,b_t,h_t)$. 
Our result shows that tax ignorance could induce more than 62\% loss on the average portfolio returns, implying the importance of tax consideration in the environment of stock trading on AI platforms. Our model could be combined with other deep learning models of stock prediction or ranking for better stock trading. 


\bibliographystyle{icml2017}
\bibliography{main.bib}

\begin{thebibliography}{8}
\providecommand{\natexlab}[1]{#1}
\providecommand{\url}[1]{\texttt{#1}}
\expandafter\ifx\csname urlstyle\endcsname\relax
  \providecommand{\doi}[1]{doi: #1}\else
  \providecommand{\doi}{doi: \begingroup \urlstyle{rm}\Url}\fi

\bibitem[David~Silver(2016)]{S16}
David~Silver, Aja~Huang, Chris J Maddison Arthur Guez Laurent Sifre George van
  den Driessche Julian Schrittwieser Ioannis Antonoglou Veda Panneershelvam
  Marc Lanctot et~al.
\newblock Mastering the game of go with deep neural networks and tree search.
\newblock \emph{Nature}, 529\penalty0 (7587):\penalty0 484--489, 2016.

\bibitem[John~Schulman \& region~policy optimization(2015)John~Schulman and
  region~policy optimization]{SM15}
John~Schulman, Sergey~Levine, Pieter Abbeel Michael~Jordan and region~policy
  optimization, Philipp Moritz.~Trust.
\newblock Trust region policy optimization.
\newblock pp.\  1889--1897, 2015.

\bibitem[John~Schulman \& Klimov(2017)John~Schulman and Klimov]{SM17}
John~Schulman, Filip~Wolski, Prafulla Dhariwal Alec~Radford and Klimov, Oleg.
\newblock Proximal policy optimization algorithms.
\newblock \emph{arXiv preprint arXiv:1707.06347}, 2017.

\bibitem[John~Schulman \& Abbeel(2016)John~Schulman and Abbeel]{SM16}
John~Schulman, Philipp~Moritz, Sergey Levine Michael I.~Jordan and Abbeel,
  Pieter.
\newblock High-dimensional continuous control using generalized advantage
  estimation.
\newblock \emph{ICLR}, 2016.

\bibitem[Min~Dai \& Fei(2015)Min~Dai and Fei]{D15}
Min~Dai, Hong~Liu, Chen~Yang and Fei, Yizhong.
\newblock Optimal tax-timing with asymmetric long-term/short-term capital gains
  tax.
\newblock volume~28, pp.\  2687--2721, 2015.

\bibitem[Sergey~Levine \& Abbeel(2016)Sergey~Levine and Abbeel]{L16}
Sergey~Levine, Chelsea~Finn, Trevor~Darrell and Abbeel, Pieter.
\newblock End-to-end training of deep visuomotor policies.
\newblock \emph{JMLR}, 17\penalty0 (39):\penalty0 1--40, 2016.

\bibitem[Volodymyr~Mnih(2015)]{M15}
Volodymyr~Mnih, Koray~Kavukcuoglu, David Silver Andrei A Rusu Joel Veness Marc
  G Bellemare Alex Graves Martin Riedmiller Andreas K Fidjeland Georg Ostrovski
  et~al.
\newblock Human-level control through deep reinforcement learning.
\newblock \emph{Nature}, 518\penalty0 (7540):\penalty0 529--533, 2015.

\bibitem[Yan~Duan \& Abbeel(2016)Yan~Duan and Abbeel]{D16}
Yan~Duan, John~Schulman, Xi Chen Peter L Bartlett Ilya~Sutskever and Abbeel,
  Pieter.
\newblock Rl: Fast reinforcement learning via slow reinforcement learning.
\newblock In \emph{NIPS Workshop on Deep Reinforcement Learning}, 2016.

\end{thebibliography}

\end{document}